\documentclass[preprint]{aastex63}

\newcommand{\sgr}{\mbox{SGR\,J1935+2154~}}
\newcommand{\sgrnos}{\mbox{SGR\,J1935+2154}}

\usepackage{amsmath}
\usepackage{amssymb}

\shorttitle{Persistent Emission Properties of \sgr}
\shortauthors{G\"o\u{g}\"u\c{s} et al.}

\graphicspath{{./}{figures/}}

\begin{document}

\title{Persistent Emission Properties of \sgr During Its 2020 Active Episode}

\correspondingauthor{E. G\"o\u{g}\"u\c{s}}
\email{ersing@sabanciuniv.edu}

\author[0000-0002-5274-6790]{Ersin G\"o\u{g}\"u\c{s}}
\affiliation{Sabanc\i~University, Faculty of Engineering and Natural Sciences, \.Istanbul 34956 Turkey}

\author[0000-0003-4433-1365]{Matthew G. Baring} \affiliation{Department of Physics and Astronomy - MS 108, Rice University, 6100 Main Street, Houston, Texas 77251-1892, USA}

\author[0000-0003-1443-593X]{Chryssa Kouveliotou} \affiliation{Department of Physics, The George Washington University, 725 21st Street NW, Washington, DC 20052, USA}\affiliation{Astronomy, Physics, and Statistics Institute of Sciences (APSIS), The George Washington University, Washington, DC 20052, USA}

\author[0000-0002-3531-9842]{Tolga G\"uver}
\affiliation{Istanbul University, Science Faculty, Department of Astronomy and Space Sciences, Beyaz\i t, 34119, Istanbul, Turkey}
\affiliation{Istanbul University Observatory Research and Application Center, Istanbul University 34119, Istanbul Turkey}

\author[0000-0002-0633-5325]{Lin Lin} \affiliation{Department of Astronomy, Beijing Normal University, Beijing 100875, China}

\author[0000-0002-7150-9061]{Oliver J. Roberts} \affiliation{Science and Technology Institute, Universities Space and Research Association, 320 Sparkman Drive, Huntsville, AL 35805, USA.}

\author[0000-0002-7991-028X]{George Younes} \affiliation{Department of Physics, The George Washington University, 725 21st Street NW, Washington, DC 20052, USA}\affiliation{Astronomy, Physics, and Statistics Institute of Sciences (APSIS), The George Washington University, Washington, DC 20052, USA}

\author[0000-0002-1861-5703]{Yuki Kaneko}\affiliation{Sabanc\i~University, Faculty of Engineering and Natural Sciences, \.Istanbul 34956 Turkey}

\author[0000-0001-9149-6707]{Alexander J. van der Horst}\affiliation{Department of Physics, The George Washington University, 725 21st Street NW, Washington, DC 20052, USA}\affiliation{Astronomy, Physics, and Statistics Institute of Sciences (APSIS), The George Washington University, Washington, DC 20052, USA}

\begin{abstract} 
We present detailed spectral and temporal characteristics of the persistent X-ray emission of \sgr based on our {\it XMM-Newton} and {\it Chandra} observations taken in the aftermath of its April 2020 burst storm, during which hundreds of energetic X-ray bursts were emitted, including one associated with an extraordinary fast radio burst. We clearly detect the pulsed X-ray emission in the {\it XMM-Newton} data.  An average spin-down rate of 1.6$\times$10$^{-11}$ s s$^{-1}$ is obtained using our spin period measurement combined with three earlier values reported from the same active episode. Our investigations of the {\it XMM-Newton} and {\it Chandra} spectra with a variety of phenomenological and physically-motivated models, concluded that the magnetic field topology of \sgr is most likely highly non-dipolar.  The spectral models indicate that surface field strengths in somewhat localized regions substantially exceed the polar value of $4.4 \times 10^{14}\,$G inferred from a spin-down torque associated with a rotating magnetic dipole.
\end{abstract}


\section{Introduction} \label{sec:intro}

Magnetars are isolated neutron stars with extremely strong magnetic fields \citep{1992ApJ...392L...9D, 1998Natur.393..235K} occasionally exhibiting unique X-ray outbursts. During the onset of such episodes, their persistent X-ray flux is elevated by 10-100 times and is usually accompanied with the emission of extremely energetic, short X-ray bursts. Their enhanced persistent X-ray intensity typically decays back to quiescence over timescales of several months to a couple of years~\citep{2018MNRAS.474..961C}. The X-ray enhancement observed during magnetar outbursts is generally attributed to dissipation of magnetic energy via the gradual untwisting of a twisted magnetosphere comprising toroidal distortions in its global dipolar field \citep{2009ApJ...703.1044B}.

Persistent X-ray emission properties of magnetars are unique, containing both thermal and non-thermal signatures over a spectral range of 0.5$-$10 keV. Their spectra can typically be described as the sum of a blackbody (BB) of $kT\sim0.5$ keV and a power law (PL) with index $\sim2$, which likely correspond to emission from the neutron star surface and the magnetosphere, respectively (see \cite{2017ARA&A..55..261K} for a review). Extensive numerical studies revealed that the emergent X-ray radiation is significantly affected by the strong magnetic fields, both throughout the neutron star atmosphere \citep[e.g.,][]{2001ApJ...563..276O,2003ApJ...583..402O,Ho-2003-MNRAS}, as well as in the magnetospheric corona \citep{Lyutikov-2006-MNRAS,Rea-2008-ApJ,2007ApJ...660..615F,Zane-2011-AdSpR}. By taking these two crucial components into account, \cite{guver07} (and later \citet{2015ApJ...805...81W}) generated physically-motivated magnetar emission models, which could yield essential magnetar characteristics through X-ray spectral modelling. Their model of the Surface Thermal Emission and Magnetospheric Scattering in 3-Dimensions (STEMS3D), has been successfully applied to the X-ray observations of magnetars in outburst \citep{guver08,guver11,gogus11,weng15}, deriving magnetar characteristics, such as the surface temperature and magnetic field strength, the magnetospheric twist angle and the average velocity of the scattering electrons in the magnetar corona \citep{beloborodov07}.

As a prolific magnetar, \sgr has emitted hundreds of short, energetic bursts in repeated outbursts since its discovery in 2014 \citep{israel16}. Based on its spin period of 3.24~s and spin-down rate of 1.43$\times$10$^{-11}$ s s$^{-1}$, \cite{israel16} determined an inferred dipole magnetic field strength, at the equatorial surface, of 2.2$\times$10$^{14}$~G. Its broadband X-ray spectral characteristics in the 2014, 2015 and 2016 outbursts are consistent with each other. Spectra from all three outbursts could be described with a BB (kT$\sim$0.5 keV) plus a PL model (index $\sim$ 2), or with two BBs with low and high temperatures of $\sim$ 0.5 and 1.6~keV, respectively \citep{israel16,younes17}. {\it NuSTAR} observations during the 2015 outburst showed that \sgr also emits X-rays up to $\sim50$\,keV \citep{younes17}. \cite{kothes18} suggested that \sgr is associated with the Galactic supernova remnant (SNR) G57.2+0.8, with an estimated distance of 12.5 kpc. However, \cite{zhou20} claim a distance to the SNR (and the magnetar) of $6.6\pm0.7$\,kpc. Independent of the SNR, \citet{mereghetti20} derived a lower distance of 4.4 kpc to \sgr. In this study, we assume a distance of 9 kpc, which is nearly the average of the highest and the lowest estimate.

On April 27$^{th}$ 2020, \sgr entered its most active episode seen to date. It emitted hundreds of short energetic bursts \citep{palmer20}, several of them bunched in clusters, over the course of a day \citep{younes20}. The source also emitted a very short radio burst, resembling an extragalactic fast radio burst \citep{chime20,stare2}, which was coincident with an X-ray burst \citep{mereghetti20,Konus20,hxmt20} with a spectrum extending beyond 100 keV. In addition to these energetic bursts, the persistent X-ray emission of \sgr went up by a factor of $\sim 55$ \citep{borghese20}. The temperature of the thermal emission component of the persistent emission was measured as 1.6~keV at the onset of the outburst, gradually cooling down to $\sim$0.5~keV in just a couple of days \citep{borghese20}. Additionally, the X-ray flux dropped to about 4-5$\times$10$^{-12}$ erg cm$^{-2}$ s$^{-1}$ in just a few days and remained nearly constant at that level for several weeks, but still about four times higher than the pre-activation flux \citep{borghese20}. During this activation, a short-lived X-ray halo around \sgr was also observed \citep{kennea20}, most likely powered by the storm of energetic bursts  \citep{mereghetti20}. 

Here, we report on the temporal and spectral investigations of \sgr using {\it XMM-Newton} and {\it Chandra} observations, taken during the flux-decay phase of its 2020 April outburst. We describe our observations and data reduction in the next section. In Section 3, we present our results and discuss their implications in Section 4.

\section{Observations and Data Reduction} \label{sec:data}

Our {\it XMM-Newton} observation of \sgr took place from 2020 May 13 21:45:43 through May 14 10:56:48, for a total exposure of nearly 50 ks. The pn-CCD instrument \citep{stueder01} of the European Photon Imaging Camera  was set to operate in the Prime Full Window mode, which affords  temporal resolution of 73 ms, allowing for spin searches as well as X-ray spectroscopy. The two Metal Oxide Semi-conductor (MOS; \citep{turnermos}) detectors operated in the same observing mode with a low temporal resolution of 2.6 s. Therefore, the data collected with MOS1 and MOS2 could only be used for spectroscopy. We used the Science Analysis System (SAS) version 18.0.0 to process the data and employed the latest calibration files to generate the response and ancillary files. We selected all events with appropriate grades (in the range from 0 to 4 for pn and 0 to 12 for MOS) but excluded those events that were registered near the edge of CCDs or hot pixels. After excluding the time intervals of higher rates due to Solar or background particle contamination, we obtained an effective exposure of 35.8 ks with the pn, 41 ks with MOS1 and with MOS2. We collected the source photons from a circular region with 30$\arcsec$ radius, which centered at RA: 19$^h$34$^m$55.$^s$59, Dec: +21$^\circ$53$'$ 47.$\arcsec$78 (J2000, \cite{israel16}). The accumulated pn, MOS1 and MOS2 spectra are grouped to contain a minimum of 50 counts per spectral bin. To form the background spectra, events were collected from circular region of 60$\arcsec$ radius source free portion on the same chip of the corresponding instrument. For timing analysis, we have also transformed the pn event arrival times to that at the Solar system barycenter using the above source position.

We also observed the source with the Advanced CCD Imaging Spectrometer (ACIS;  \citet{townsley}) onboard the {\it Chandra} X-ray Observatory starting at 2020 May 18 10:48:14 for 20 ks. The {\it Chandra} observation was taken with the nominal 3.14 s frame readout time, which is too coarse to probe the 3.24 s spin period of the source, therefore, we only performed phase-averaged X-ray spectroscopy of the magnetar. We used {\it Chandra} Interactive Analysis of Observations version 4.12 and CALDB 4.9.2.1 to process the data and generate the calibration files. 
We extracted photons within a circle of 6$\arcsec$ radius to generate the source spectrum. The background spectrum was formed using events within a circular source free region of 40$\arcsec$ radius.

\section{Data Analysis and Results}

\subsection{Temporal Analysis} \label{sec:tempo}

We performed a period search in the {\it XMM-Newton} data using the unbinned events, with the Z$^2_m$ statistic with {\it m} = 2 harmonics \citep{buccheri} in an interval from 3.2473 to 3.2474 s. We found a significant signal at the frequency of 0.3079437(4)\footnote{The reported errors are at a 1$\sigma$ confidence level throughout the paper.} Hz, which corresponds to a spin period of 3.247345(4) s. Recently \citet{borghese20} reported spin period measurements of 3.24731(1) s based on {\it NICER} observations on 2020 April 29-30, and of 3.247331(3) s and 3.24734(1) s, based on {\it NuSTAR} observations on 2020 May 2 and May 9$-$10, respectively. We then combined our {\it XMM-Newton} spin period measurement with these three, and fit a first order polynomial to the four spin measurements spanning from 2020 April 29 through May 14, to obtain an average spin-down rate of 1.6(5)$\times$10$^{-11}$ s s$^{-1}$.

\begin{figure*}
\begin{center}
\includegraphics[scale=0.6]{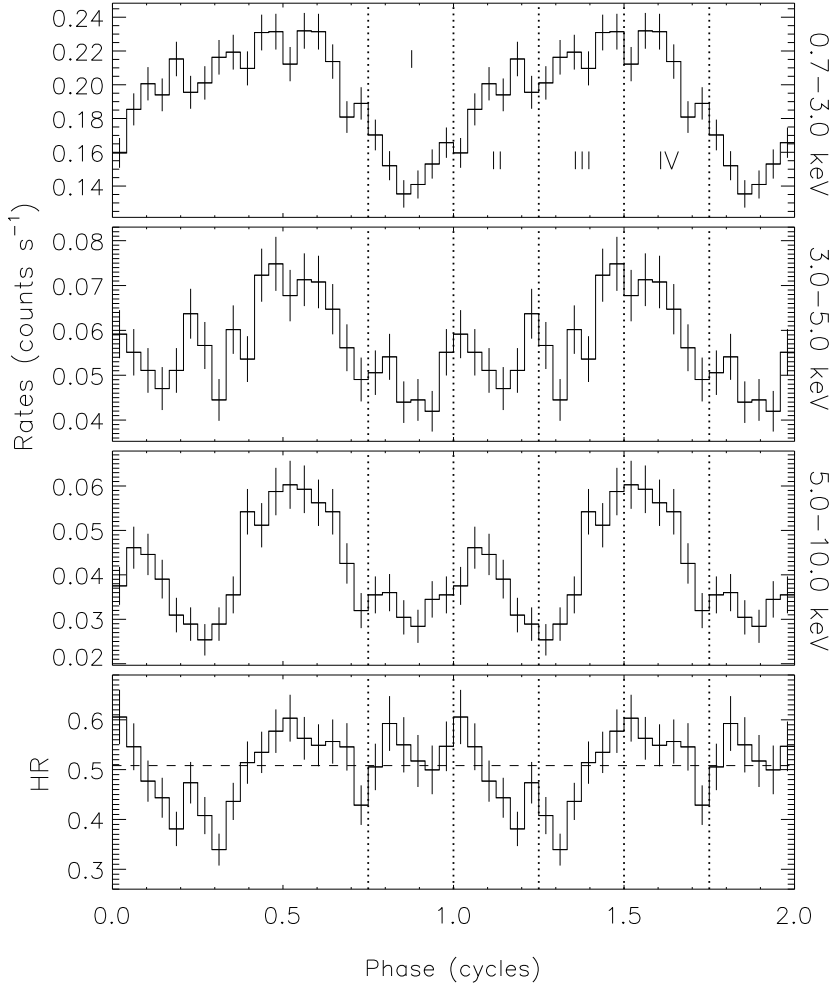}
\end{center}
\vspace{-2.0cm}
\caption{{\it (Upper three panels)} Energy dependent pulse profiles of \sgr folded with the {\it XMM-Newton} spin period. The corresponding energy interval of each profile is indicated on the right y-axis. The vertical dotted lines with I-IV indicate the phase intervals used in phase-resolved spectral analysis. {\it (Bottom panel)} The ratio of spin phase dependent hard count rates in the 3-10\,keV band to those in the 0.7-3.0\,keV. The dashed horizontal line denotes the mean hardness ratio of 0.51. \label{fig:pulse}}
\end{figure*}

We have constructed the energy resolved pulse profiles of \sgr based on our spin period measurement using {\it XMM-Newton}. In the upper three panels of Figure \ref{fig:pulse}, we present pulse profiles over energy bands of 0.7-3, 3-5, and 5-10\,keV. We find that the pulse profile in the lowest energy band is dominated by a broad structure with an RMS pulsed fraction of $0.14(1)$. A similar RMS pulsed fraction of 0.14(2) is observed in the next band (3-5 keV), but with substructures becoming more prominent in the profile. The 5-10\,keV band has a significantly higher RMS pulsed fraction of 0.25(2), and is dominated by a main structure spanning about 35\% of the phase, with a second pulse of similar width but less prominent. 

The bottom panel of Figure \ref{fig:pulse} is the hardness ratios over the phase intervals. Here we defined this measure as the ratio of the phase dependent count rates in the 3-10\,keV (i.e., the sum of two middle panels) to those in the 0.7-3.0\,keV. We find that the RMS amplitude of the hardness ratios differs by 0.26(7) with respect to their mean value of 0.51 (marked with the dashed horizontal line in Figure \ref{fig:pulse}).

% fig: pulse profiles

\subsection{Spectral Analysis} \label{sec:spectral}

\begin{figure*}
\begin{center}
\includegraphics[scale=0.7,angle=180]{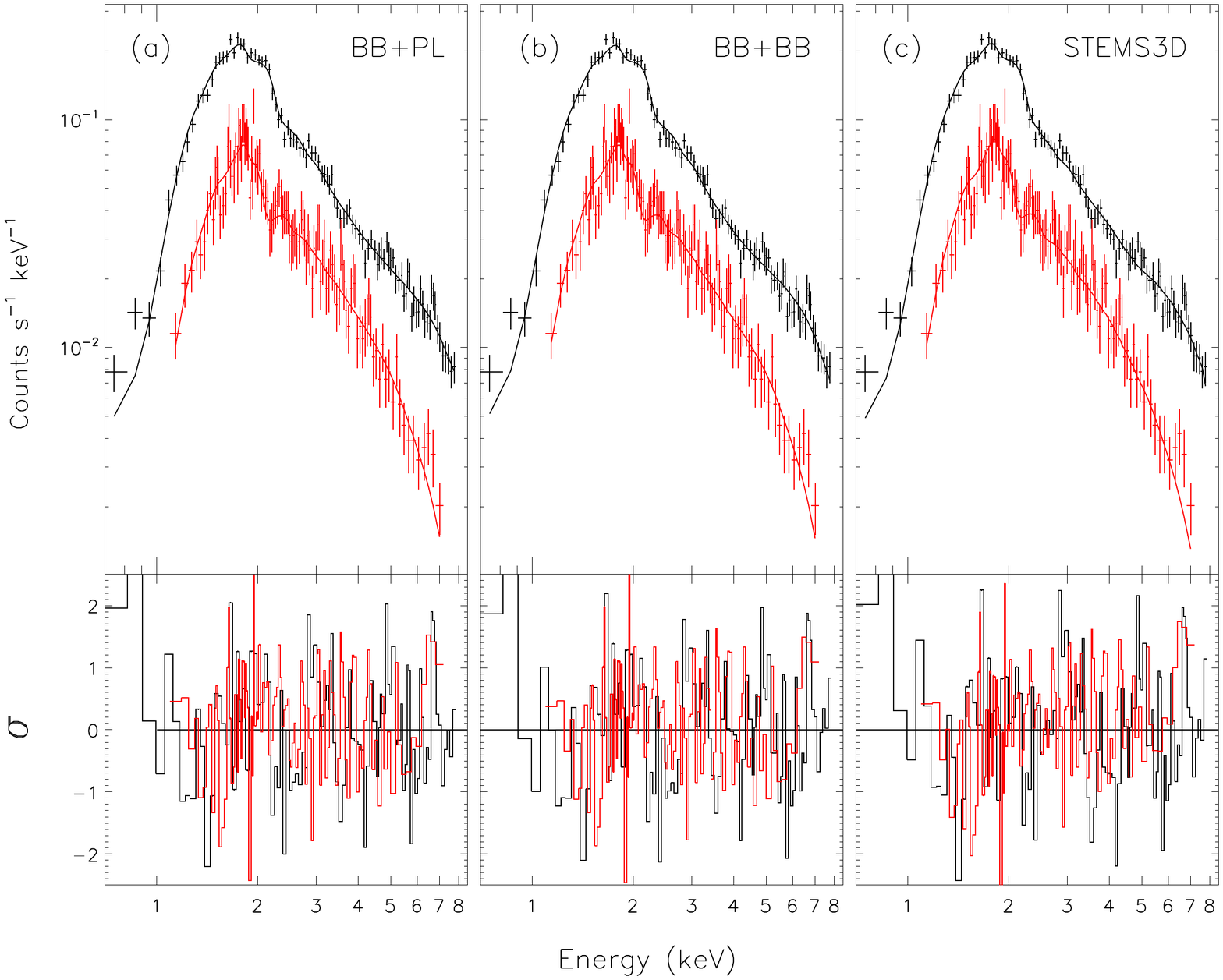}
\end{center}
\vspace{-1.0cm}
\caption{Phase-averaged spectral energy distribution of \sgr detected with {\it XMM-Newton} {EPIC-pn} in black and {\it Chandra} in red. For clarity, only these two spectra are displayed in each panel since MOS1 and MOS2 spectra appear overlapped with {\it Chandra}. (a) Best fit BB+PL model with fit residuals. (b) Best fit BB+BB model with residuals. (c) Best fit STEMS3D model with residuals. 
\label{fig:spec}}
\end{figure*}

We performed joint spectral fits of the {\it XMM-Newton} (i.e., pn, MOS1 and MOS2) and {\it Chandra} spectra in the 0.7$-$8.5 keV energy range. More specifically, we linked the model parameters for both data sets to yield a better constrained value, while allowing their amplitudes (normalizations) to vary independently for each set. Initially, we fit a BB + PL model, which phenomenologically describes magnetar spectra adequately. To account for the effects of interstellar absorption, we used the model by \cite{wilms} in XSPEC \citep{arnaud}. We obtain a good fit to both spectra, with a $\chi^2$ of 357.2 for 346 degrees of freedom (dof). The corresponding model parameters are: the interstellar hydrogen column density, $N_{\rm H}$ = (2.6$\pm$0.2)$\times$10$^{22}$ cm$^{-2}$, the BB temperature, $kT=0.41\pm$0.02 keV, and the PL index, $\Gamma = 1.4\pm 0.2$. We note that this index is consistent with the value of $\Gamma \approx 1.2$ obtained  for the energy range $\sim$4$-$20 keV by \cite{borghese20} on May $2^{nd}$ and $11^{th}$, 2020 via {\sl NuSTAR} observations. We present the EPIC-pn and {\it Chandra} as two representative spectra with the best fit BB and PL parameters in the panel (a) of Figure \ref{fig:spec}. The 0.5$-$10 keV unabsorbed X-ray flux with {\it XMM-Newton} is (5.3$\pm$0.2)$\times$10$^{-12}$ erg s$^{-1}$ cm$^{-2}$, while the {\it Chandra} unabsorbed flux, measured five days later, is slightly lower, (4.4$\pm$0.2)$\times$10$^{-12}$ erg s$^{-1}$ cm$^{-2}$. The corresponding isotropic luminosities for {\it XMM-Newton} and {\it Chandra} are 5.1 $\times$10$^{34}$ erg s$^{-1}$ and 4.3$\times$10$^{34}$ erg s$^{-1}$, respectively.

% fig: allmodels_bbbbline

Modelling the four spectra jointly with the sum of two BB functions yields similar fit statistics ($\chi^2$ of 362.6 for 346 dof). We find an $N_{\rm H}$ of 2.4$\pm$0.2$\times$10$^{22}$ cm$^{-2}$ and BB temperatures of $0.44\pm0.02$\, keV and $1.9\pm0.2$\,keV (see panel (b) in Figure \ref{fig:spec}). The BB amplitudes yield radii for the emitting regions of 3.1$^{+0.6}_{-0.9}$\,km and 0.13$\pm$0.03\,km for the cool and the hot BB components, respectively. Accordingly, the cooler BB may originate from a sizable fraction of the stellar surface, a deduction commensurate with the broad peak in the pulse profile below $\sim 3\,$keV, as displayed in Figure~1.

We then fit the X-ray spectra jointly with the physically motivated STEMS3D model. This model also provides statistically acceptable fits to all four spectra ($\chi^2=370.1$ for 348 dof). We find that $N_{\rm H}$ is (2.5$\pm$0.2)$\times$10$^{22}$ cm$^{-2}$, the surface temperature of the neutron star is $0.52\pm0.03$\,keV, and the surface magnetic field is ($9.6\pm0.2)\times10^{14}$ G. The magnetospheric components of the STEMS3D yield a twist angle of 0.29$\pm$0.04 and an average electron speed of ($0.59\pm0.05$)c, where c is the speed of light. In panel (c) of Figure \ref{fig:spec}, we present the two representative spectra with the best fitting STEMS3D model parameters.   

To explore any phase dependent spectral variations we performed spin phase-resolved spectroscopy with the BB+PL, BB+BB and STEMS3D models. We accumulated four {\it XMM-Newton}-pn spectra from the phase intervals indicated as I, II, III and IV in Figure \ref{fig:pulse}. We allowed the parameters of these continuum models to vary across phases and linked the interstellar H absorption to yield a common value for each. We find in all three cases that the resulting N$_H$ values are consistent with those of the phase-integrated spectroscopy. For the BB+PL model, we find that the BB temperature remains nearly constant (within errors) around 0.39 keV but the power law index indicates a slight variation (see Table \ref{phspectab}). For the BB+BB model, the temperature of the cooler BB component remains around 0.42 keV, while the temperature of the hot BB component varies marginally between 1.5 keV and 2.3 keV (Table \ref{phspectab}). For the STEMS3D modelling, we have linked the magnetospheric electron velocity parameter across phases to better constrain the other model parameters. The surface temperature remains consistent around 0.54 keV, while the surface magnetic field strength hints to a marginal increase in phase interval IV. A final element to note is that in Fig.~\ref{fig:spec} there appears to be a modest excess or emission feature in a narrow energy range centered on around 6.7 keV in both {\it XMM-Newton} and {\it Chandra} spectra.  Our detailed data analysis and simulations indicate that it is marginal, with significance in the $1.5-3\sigma$ range in phase intervals II-IV. 

\begin{deluxetable*}{ccc|cc|cccc}
\tabletypesize{\small}
\tablecaption{Results of Phase-Resolved Spectral Analysis Using EPIC-pn Data
  \label{phspectab}
}
\tablewidth{\textwidth}
\tablehead{
\colhead{}  & \multicolumn{2}{c}{BB+PL} & \multicolumn{2}{c}{BB+BB} & \multicolumn{4}{c}{STEMS3D} \\
\colhead{Phase} & \colhead{kT} & \colhead{$\Gamma$} & \colhead{kT$_c$} & \colhead{kT$_h$} & \colhead{kT} & \colhead{B} & \colhead{Twist} & \colhead{$\beta$} \\
 & (keV) & & (keV) & (keV) & (keV) & (10$^{14}$ G) & angle & (c) }

\startdata
I   & 0.36$\pm$0.02 & 0.9$\pm$0.5 & 0.39$\pm$0.03 & 1.5$\pm$0.3 & 0.57$\pm$0.07 & 8.8$\pm$0.5 & 0.24$\pm$0.04 & 0.67$\pm$0.09  \\      
II  & 0.41$\pm$0.02 & 1.7$\pm$0.3 & 0.45$\pm$0.03 & 2.3$\pm$0.4 & 0.57$\pm$0.08 & 8.5$\pm$0.3 & 0.21$\pm$0.04 & (L)  \\      
III & 0.39$\pm$0.02 & 1.3$\pm$0.3 & 0.43$\pm$0.03 & 1.7$\pm$0.3 & 0.58$\pm$0.07 & 8.5$\pm$0.3 & 0.23$\pm$0.04 & (L)   \\      
IV  & 0.39$\pm$0.02 & 1.9$\pm$0.3 & 0.43$\pm$0.03 & 1.9$\pm$0.3 & 0.51$\pm$0.06 & 9.8$\pm$0.3 & 0.29$\pm$0.05 & (L)   \\
\hline
\enddata
\end{deluxetable*}

\section{Discussion}

The recent active episode of \sgr has been very prolific: besides numerous energetic bursts and a burst storm, a fast radio burst (FRB) associated with an X-ray burst was also detected, the first from a Galactic magnetar. Recently \cite{borghese20} reported three spin period measurements of \sgr enabling them to place a 3$\sigma$ upper limit to the spin down rate, $\dot{P}$, of 3$\times$10$^{-11}$ s s$^{-1}$. In this study, we combined our spin period measurements with {\it XMM-Newton} on 2020 May 13-14 with the 
timing data from \cite{borghese20} and were able to obtain an average $\dot{P}$ of 1.6$\times$10$^{-11}$ s s$^{-1}$ for the time interval spanning from 2 to 17 days after its reactivation. This is in agreement with the spin-down rate measured with observations performed from 2014 July through 2015 March. \sgr rotational dynamics, therefore, follow the prevailing magnetar trend that burst active periods (including storms) do not impose significant immediate torque changes on the neutron star, while torque changes have been observed prior and after multiple bursts and giant flares \citep[e.g.,][]{Younes-2017-ApJ-1806}.  Interestingly, using {\it NICER} observations of \sgr collected within the time span of 21 to 39 days after the latest outburst onset (in the absence of intense bursts), \cite{younes20} precisely measured a much higher spin-down rate of 3.9$\times$10$^{-11}$ s s$^{-1}$. This might suggest a hysteresis in the dynamical response in these highly magnetized systems that is accompanied by changes in burst activation.

The polar magnetic field strength inferred for a vacuum dipole approximation (i.e., B$_p = 6.4\times10^{19}~G~\sqrt{P\dot{P}}\;$) of \sgr is about 4.4$\times10^{14}$ G. However, our X-ray spectral modelling of the persistent emission with the STEMS3D model, which includes radiative propagation in extremely magnetized settings, yielded a local surface magnetic field strength of 9.6$\times10^{14}$ G.  This difference is not surprising since the spin-down estimate applies to the global field configuration, and can be modified by plasma loading of the magnetosphere \citep{Harding-1999-ApJ}. Nevertheless, the surface magnetic field strength being considerably larger than the inferred value suggests that there may be multi-polar magnetic topology, where local field strengths are much larger than the dipolar ones. Such a circumstance was observed in SGR 0418+5729, the first `low-dipole field' magnetar. Its magnetic field strength inferred from its spin parameters is 6$\times10^{12}$ G \citep{rea10}. \cite{guver11} modelled the {\it XMM-Newton} spectrum of SGR 0418$-$5729 with the STEMS atmospheric emission model and obtained $\sim 10^{14}$ G for the surface magnetic field strength. They proposed a dominant non-dipolar magnetic field structure for SGR 0418$-$5729, a picture that is consistent with the discovery of a variable proton cyclotron absorption feature in later {\it XMM-Newton} observations \citep{tiengo13}. 

A significant multi-polar magnetic configuration for \sgr is also supported by the complexity of its energy dependent pulse profiles. Its low energy X-ray pulse profile is dominated by a broad structure with a phase width of around $0.7-0.8$. However, minor structures become pronounced above $\sim$3 keV, in the absence of bright persistent X-ray emission ($kT=0.5$\,keV) from the surface of the neutron star.  The BB+BB spectral fits indicate that this putatively hotter portion emanates from smaller regions, less than $\sim0.2$\,km in size, and perhaps analogous to multi-polar field components that are observed from small regions on the Sun.  These structured pulse profiles contrast the comparatively simple two-peaked forms observed for active outburst phases in another magnetar, 1RXS J170849.0$-$400910, which can be attributed to antipodal polar hot zones in a dipole field configuration \citep{Younes-2020-ApJ-1708}. Extension of the pulse profile analysis of \cite{Younes-2020-ApJ-1708} to treat broader surface temperature profiles in magnetic colatitude is unlikely to generate the abrupt rises and falls in consecutive phase bins evident above $\sim 3\,$keV in Figure~1.  We suggest that strong temperature gradients in magnetic longitude are also required to reproduce the observed profiles, and these are naturally generated in multi-polar magnetic morphologies. 

We found that there is no significant spin phase dependent spectral variations for any of the three continuum models employed. The results of the phase resolved spectral analysis using EPIC-pn data only are consistent (within 2$\sigma$ level) with the model parameters obtained with the joint analysis of the four spectra. Investigations with the STEMS3D model (both phase-resolved or phase-averaged) reveal that the magnetospheric twist angle, $\Delta\phi\sim 0.3$ in the aftermath of its major outburst is modest, and the average magnetospheric electron speed ($\beta\sim$ 0.6) is moderate. \citet{2015ApJ...805...81W} applied the STEMS3D model to a large number of {\it XMM-Newton} observations of magnetars in outbursts. They found that the magnetosphere of the variable magnetars are highly twisted ($\Delta\phi$ $>$ 1). The modest magnetospheric twist in \sgr implies that the energy budget supplied via dissipation through untwisting would be low \citep{2009ApJ...703.1044B}. This was, in fact, the case for the 2020 outburst of \sgr as its persistent X-ray emission enhancement was mild and decayed more than an order of magnitude in just few days following the onset \citep{borghese20,younes20}.

\acknowledgments
We thank the anonymous reviewer for helpful comments. We are grateful to Norbert Schartel and Belinda Wilkes for approving {\it XMM-Newton} and {\it Chandra} DDT observations of \sgrnos. M.G.B. acknowledges the generous support of the National Science Foundation through grant AST-1813649.  C.K. acknowledges support by the Smithsonian Astrophysics Observatory under grant DDO-21120X. L.~L. acknowledges support from the National Natural Science Foundation of China (grant number 11703002).

\bibliographystyle{unsrt}
\bibliography{refs.bib} 

\end{document}